**Pulsed Cathodoluminescence Spectra of Solid Oxides with Low Concentrations of Optically-Active Impurities**

*V.A. Kozlov* [a], *S.A. Kutovoi* [b], *N.V. Pestovskii* [a,c], *A.A. Petrov* [a,c], *S.Yu. Savinov* [a,c], *Yu.D. Zavartsev* [b], *M.V. Zavertyaev* [a] *and A.I. Zagumenniy* [b]

[a]P.N. Lebedev Physical Institute of the Russian Academy of Sciences, 119991, 53 Leninskiy Prospekt, Moscow, Russian Federation.

[b]A.M. Prokhorov Institute of General Physics of the Russian Academy of Sciences, 119991, 38 Vavilov Str., Moscow, Russian Federation.

[c]Moscow Institute of Physics and Technology (State University), 141700, 9 Institutskiy per., Dolgoprudny, Moscow Region, Russian Federation.

*Pulsed cathodoluminescence (PCL) spectra of ultra-pure $SiO_2$, $GeO_2$, $SnO_2$, $TiO_2$, $La_2O_3$, $Y_2O_3$, $Sc_2O_3$, $CaCO_3$ powders and α-quartz, Ca:$YVO_4$, $LiNbO_3$ and Sc:$LiNbO_3$ crystals were studied under the same experimental conditions. It was found that PCL spectra of $SiO_2$, $SnO_2$, $GeO_2$, $TiO_2$, $La_2O_3$ and $CaCO_3$ powders contain a common band with maximum intensity at 500 nm, PCL spectra of samples $Y_2O_3$, $Sc_2O_3$, $PbWO_4$ and Ca:$YVO_4$ contain a common band at 490 nm and PCL spectra of $LiNbO_3$ and Sc:$LiNbO_3$ crystals contain a common band at 507 nm. It was found that the average intensity of the PCL spectra and position of the maximum intensity of these common bands depend on the type of a band gap transition of the material. We suppose that these common bands have the same origin in PCL spectra of all the materials studied and are related to recombination of $O^{2-}$-$O^-$ oxygen complexes. These complexes appear in the vicinities of anionic and cationic vacancies, where the geometry and orientation of coordination polyhedrons are violated due to the stress in the lattice.*

**Corresponding author:** Nikolai V. Pestovskii, e-mail:pestovsky@phystech.edu

**Highlights:**

- Pulsed cathodoluminescence (PCL) spectra of ultra-pure $SiO_2$, $GeO_2$, $SnO_2$, $TiO_2$, $La_2O_3$, $Y_2O_3$, $Sc_2O_3$, $CaCO_3$ powders and α-quartz, Ca:$YVO_4$, $LiNbO_3$ and Sc:$LiNbO_3$ crystals with low concentration of optically-active impurities were studied under the same experimental conditions.

- A common band in PCL spectra of all samples is found. PCL spectra of $SiO_2$, $GeO_2$, $La_2O_3$, $SnO_2$, $TiO_2$ and $CaCO_3$ powders possess a common band at 500 nm. PCL spectra of α-quartz, Ca:$YVO_4$ and $PbWO_4$ crystals and $Y_2O_3$ and $Sc_2O_3$ powders possess a common band at 490 nm, and the PCL spectra of $LiNbO_3$ and Sc:$LiNbO_3$ crystal possess a common band at 507 nm.

- It was found that the position of these bands in PCL spectra and the average intensity of PCL spectra of the sample are related to the type of band gap transition.

- We attribute this band to luminescence of $O^{2-}$-$O^{-}$ oxygen complexes in lattice violation regions neighboring the anionic and cationic vacancies.

## 1. Introduction

Solid oxides are widely used in all fields of science and technology. Materials with low concentrations of impurities are of prime importance in many applications. For instance, these substances are widely used as starting materials for crystal growth or in drug manufacture, for production of micro and nanoelectronic devices, for creation of the optical communication lines and spintronic systems, and in other applications.

Luminescence analysis is a sensitive and non-destructive method of controlling the impurity concentration in substances [1]. Establishing the luminescence properties of ultra-pure materials is necessary to define their luminescent markers. Using these markers, it is possible to examine the chemical purity of samples and the defect concentration in their structure.

The purpose of this study is to measure the luminescence spectra of a wide range of solid oxides, containing low concentrations of optically-active impurities, under the same experimental conditions for all samples, and to establish the common features of their luminescence spectra.

Materials with different properties (chemical composition, level of structural regularity, crystal structure, type of chemical bond, etc.) were studied.

$SiO_2$, $GeO_2$, $SnO_2$, $TiO_2$, $Y_2O_3$, $Sc_2O_3$, $CaCO_3$, $PbWO_4$, $YVO_4$ and $LiNbO_3$ materials are related to a group of insulators or the wide-band gap semiconductors. Luminescence of $SiO_2$ had been studied in [2-26]. A review of optically-active defects in $SiO_2$ is given in [26]. The blue band of $SiO_2$ in the region of 2.7-3 eV (410-460 nm) is attributed to the Self-Trapped Exciton (STE) recombination in [2-4,6,8,10-13,16], in [26] it is also attributed to luminescence of Oxygen-Deficiency Centers (ODC), and in [23] it is attributed to luminescence of $Al^{3+}$ impurity.

Luminescence at 1.9 eV (653 nm) of amorphous $SiO_2$ was studied in [4,7,14,15,18-20,26] and attributed to the recombination of the Non-Bridging Oxygen-Hole Centre (NBOHC). This band does not arise in the luminescence spectra of non-irradiated crystalline quartz [26]. A review of different luminescence bands in the spectrum of $SiO_2$ was also given in [16]. Photoluminescence (PL) spectra of silica nanoparticle composites with particle size of 7 and 15 nm had been studied in [18-20]. PL spectra [18-20] consisted of two bands with maxima at 2.37 eV (523 nm) and at 2.7 eV (460 nm). The structure of the band at 2.37 eV consisted of three peaks in the spectrum of 7 nm powder and four peaks in the spectrum of 15 nm powder. The authors attributed a band at 2.7 eV to STE and band at 2.37 eV to luminescence of OH radicals, adsorbed on the nanoparticle surface. The PL of $SiO_2$ nanopowder with maxima at 3.0 eV, 2.6 eV and 2.7 eV was attributed in [22] to transitions in nanosilicon network, embedded within the oxide.

Luminescence of $GeO_2$ had been studied in [21,24]. Luminescence of $SnO_2$ powder had been studied in [27,28]. In [27] it was related to oxygen vacancies. In [28] it was attributed to electronic transitions from the bottom of the conduction band and from the energy levels of oxygen vacancies to the levels of surface states. Luminescence of $Y_2O_3$ had been studied in [29-31]. The band at 3.5 eV (354 nm) is attributed in [29,30] to STE luminescence. Luminescence of $Sc_2O_3$ was studied in [32]. The band at 3.5 eV was attributed to STE similar to $Y_2O_3$. Other bands in [32] at 3.05 eV (406 nm), 2.65 eV (468 nm), 2.35 eV (528 nm) and 2.05 eV (605 nm) were attributed to luminescence of defect centers.

Luminescence of substances with extremely low concentration of optically-active impurities in the transparency region can be caused by the exciton recombination or by the recombination of centers associated with different defects in the volume or at the surface of the sample [33]. Excitation with high energy density is necessary to pump the luminescence of ultra-pure transparent samples. Especially, it is required to study the luminescence of ultra-pure substances at room temperature, when the temperature quenching of luminescence is extremely significant. At the same time, the possibility of observing luminescence at room temperature is of prime importance for applications.

In the present work the room-temperature pulsed cathodoluminescence (PCL) spectra of a set of solid oxides with different properties and low concentrations of optically-active impurities are studied. The effect of crystalline order, chemical composition and type of transition at the fundamental absorption edge is analyzed. The common features of the luminescence of solid oxides are discussed using the spectral data obtained.

## 2. Experimental setup

The PCL method of studying the condensed substances [34-39] is based on the analysis of luminescence emitted by a sample after irradiation by a short (~2 ns) and high-power (~10 $MW/cm^2$) electron beam with average energy of particles ~150 keV. It was shown [37-39] that PCL spectra and decay times of the crystals are identical to the spectra and decay times of the luminescence of the same crystals excited by gamma-irradiation, – gamma-luminescence (GL). At the same time, the PCL intensity is higher than GL intensity by several orders of magnitude.

High energy density of incident electron beam makes it possible to trace the existence of impurities and intrinsic defects in sample with a high accuracy and allows the luminescence to be captured even from opaque samples. Thus the PCL method provides a possibility of studying the luminescence from ultra-pure transparent crystal matrices and ultra-pure powders with 99.9999% purity.

The PCL spectra of all samples were studied under the same experimental conditions. This means that all spectra were recorded using the same experimental setup at room temperature. This fact makes the correct comparison of spectra of different samples possible. The experimental setup was similar to [37-39]. The luminescence from the sample was transported into the entrance slit of an OCEAN USB2000 spectrometer via the quartz optical fiber. The fiber was used to carry out an optical decoupling of the accelerator and spectrometer. The resolution of the spectrometer was 1.5 nm, and its spectral range was 200-800 nm. All spectra were corrected in accordance with the spectral sensitivity of the optical system. The CCD-matrix of spectrometer accumulated the light signal during 30-sec exposure. The repetition rate of electron pulses of the RADAN-EXPERT accelerator [35,36] was 1 Hz. Thus the measured PCL spectra are the results of the sum of intensities of all light signals emitted within 30 sec. It means that the measured spectra are the results of an average of 29 pulses of the accelerator.

The penetration depth of the electron beam with average electron energy ~150 keV into the solid sample is about 0.05-0.5 mm depending on the density and average atomic number of the material [40]. The cross-sectional area of the electron beam of the accelerator is ~3 $cm^2$. At all points of the cross-section the peak energy density is about ~10 $MW/cm^2$. Thus it is possible to capture the luminescence from a large area of the sample, maintaining a high energy density at all points of the cross-section. This is impossible in PL, because such a high energy density can only be obtained using a highly-focused light beam [18-20]. In this case the volume and cross-section of the emitting medium is much smaller than in the case of PCL.

In the space between the output window of the IMAZ-3E electron tube of the RADAN-EXPERT electron accelerator and the sample under investigation the emission of second positive system of bands of molecular nitrogen $N_2$ (SPS) and first negative system of bands of molecular ion $N_2^+$ (FNS) occurs. This space is filled with ambient air and has a length of ~3 cm.

The radiation spectra during the excitation of air by the electron beam in the absence of a sample are shown in Fig 1. Peaks of intensity at 316 nm, 337 nm, 358 nm, 367 nm, 376 nm, 380 nm, 391 nm, 400 nm and 406 nm correspond to electron-vibrational bands with unresolved rotational structure of $C^3\Pi_u$-$B^3\Pi_g$ transition (SPS) of molecular $N_2$. The peak at 426 nm corresponds to the (0.1) electron-vibrational band of $B^2\Sigma_u^+$-$X^2\Sigma_g$ transition of molecular ion $N_2^+$ [41]. This system of bands exists in all the measured spectra. If the intensity of luminescence of the sample is comparable to the intensity of SPS and FNS emission, FNS and SPS bands become visible in PCL spectra.

The intensity of SPS and FNS bands in PCL spectra is proportional to the total energy of electron irradiation adsorbed by the sample. This fact allows the proper comparison of intensities of spectra of different samples despite the pulse energy variation of the RADAN-EXPERT accelerator from pulse to pulse [35].

## 3. Samples

Ultra-pure $SiO_2$, $GeO_2$, $SnO_2$, $TiO_2$, $La_2O_3$, $Y_2O_3$, $Sc_2O_3$ and $CaCO_3$ powders had a chemical purity of at least 99.99%. The particle size of all the powders was in the order of 1 μm. The $La_2O_3$ sample had been annealed at 1100°C before the measurement of PCL spectra to decrease the concentration of water molecules and OH-groups on the surface of the powder particles. Packages with all other powders were opened immediately before the PCL measurement to prevent the adsorption of water vapour or other contaminations on their surfaces.

A single crystal of α-quartz had been grown using the hydrothermal method and had an extremely low concentration of impurities. Ca:$YVO_4$, $PbWO_4$, $LiNbO_4$ and Sc:$LiNbO_4$ crystals had been grown using the Czochralski (CZ) technique by means of the inductive heating of an iridium crucible. The purity of the starting materials was at least 99.99%. The $LiNbO_4$ and Sc:$LiNbO_4$ samples had been grown from a congruent melt composition [42].

The $Ca^{2+}$ and $Sc^{3+}$ ions doped into the $LiNbO_4$ and $YVO_4$ matrices are not optically active. However, they dominate the defect concentration [42-44]. Thus, no changes were found in the refractive index of Sc:$LiNbO_3$ in dependence on the $Sc^{3+}$ concentration and only a small change in UV-absorption was observed [42].

## 4. Results

The PCL spectra of $SiO_2$, $GeO_2$, $TiO_2$, $La_2O_3$ and $CaCO_3$ ultra-pure powders are shown in Fig. 2. The common narrow lines in the spectra of all the powders in the range of 300-410 nm correspond to the emission of SPS and FNS of molecular nitrogen $N_2$ and molecular ion $N_2^+$ (Fig. 1).

Fig. 2 shows that the spectra consist of at least two broad bands. One of these bands is common for spectra of all materials and has a maximum of intensity at 500 nm. It is clear from Fig. 2 and from Table 1 that the chemical composition, stoichiometric ratio and the percentage of ionicity of chemical bond (for example, 22% for $SiO_2$ and 67% for $La_2O_3$ [45]) has no effect on the position of the band. Band gap width and type of coordination polyhedron also have no effect on the position of maximum of intensity.

In Table 1 the corresponding characteristics of each sample are shown. It can be seen from Table 1 however, that the common feature of $SiO_2$, $GeO_2$, $TiO_2$ and $CaCO_3$ materials is an indirect band gap. Fig. 2 shows that SPS lines have a comparable intensity with the luminescence of $SiO_2$, $GeO_2$, $TiO_2$, $La_2O_3$ and $CaCO_3$ powders. This fact is due to the weak luminescence of these materials. We also attribute this fact to the indirect character of band gap transition.

The position of another band in the PCL spectra of $SiO_2$, $GeO_2$, $TiO_2$, $La_2O_3$ and $CaCO_3$ is at 430 nm (2.9 eV) for $SiO_2$, $GeO_2$, $TiO_2$, $La_2O_3$ and at 410 nm (3 eV) for $CaCO_3$. In addition to these two bands the PCL spectrum of $SiO_2$ contains bands in the region of 635 nm (1.95 eV) and in the region of 300-325 nm (3.8-4.1 eV). Absence of $SiO_2$ emission at wavelengths shorter than 250 nm can be explained by the absorption of UV-radiation by the quartz fiber. The PCL spectrum of $La_2O_3$ contains three narrow lines at 495 nm, 511 nm and 668 nm. We attribute these lines to uncontrolled rare earth impurities. Also the PCL spectrum of $La_2O_3$ contains a broad band with a maximum of intensity at 630 nm (1.97 eV). It is seen from Fig. 2 that the band at 500 nm (2.48 eV) is dominant in the PCL spectra of $SiO_2$, $GeO_2$ and $La_2O_3$. In the PCL spectra of $TiO_2$, and $CaCO_3$ intensities of bands at 500 nm and at 400-450 nm are equal.

Fig. 3 shows the PCL spectra of α-quartz, Ca:$YVO_4$, $PbWO_4$, $LiNbO_3$ and Sc:$LiNbO_3$ crystals. All crystals contained a low concentration of optically active impurities. The PCL spectra of α-quartz, Ca:$YVO_4$, $PbWO_4$, $LiNbO_3$ and Sc:$LiNbO_3$ crystals consist of at least two wide bands with flat maxima similar to Fig. 2. The PCL spectrum of an α-quartz single crystal is composed of two wide bands with maxima of intensity at 490 nm (2.53 eV) and in the region of 410 nm (3.02 eV). The band at 490 nm (2.53 eV) possesses a structure with a step at 523 nm (2.37 eV). PCL bands with maxima at 490 nm (2.53 eV) are also observed in the spectra of Ca:$YVO_4$ and $PbWO_4$ crystals. The PCL spectra of Ca:$YVO_4$ and $PbWO_4$ consist of two bands. The maximum of PCL intensity of another band in the spectrum of Ca:$YVO_4$ is situated in the region of 440 nm (2.82 eV) and in the spectrum of $PbWO_4$ one is situated in the region of 430 nm (2.88 eV).

The PCL spectrum of Sc:$LiNbO_3$ consists of four bands with maxima at 420 nm (2.95 eV), 507 nm (2.45 eV) and 765 nm (1.62 eV). The band with maximum of intensity at 507 nm (2.45 eV) possesses a step at 635 nm (1.95 eV). The PCL spectrum of $LiNbO_3$ consists of three wide bands with maxima at 410 nm (3.02 eV), 507 nm (2.45 eV) and 635 nm (1.95 eV). Consequently, the PCL spectra of $LiNbO_3$ and Sc:$LiNbO_3$ possess a common band at 507 nm (2.45 eV). Fig. 3 illustrates that the band with maxima in the region of 635 nm (1.95 eV) in the PCL spectrum of $LiNbO_3$ is situated at the same wavelength, as a step on the Sc:$LiNbO_3$ PCL spectrum.

The PCL spectra of $Y_2O_3$ and $Sc_2O_3$ ultra-pure powders are shown in Fig. 4. The spectra of these two samples also consist of two wide bands centered at similar wavelengths. However, intensity distribution in the bands is different in comparison with the case of Fig. 2. The spectra of $Y_2O_3$ and $Sc_2O_3$ ultra-pure powders exhibit luminescence with maxima at 355 nm (3.49 eV) and 488 nm (2.54 eV) for both samples. Narrow lines in the range of 550-700 nm in the PCL spectrum of $Sc_2O_3$ we attribute to $Eu^{3+}$ impurity. The average intensity of the PCL spectra of $Y_2O_3$ and $Sc_2O_3$ powders is higher than the average intensity of the PCL spectra of $SiO_2$, $GeO_2$, $TiO_2$, $La_2O_3$ and $CaCO_3$. Also, the position of the common band is shifted to 490 nm (2.53 eV) and the distribution of intensity between bands in the spectra is different. In contrast to samples from Fig. 2, $Y_2O_3$ and $Sc_2O_3$ possess a direct band gap transition (Table 1).

The PCL spectrum of ultra-pure $SnO_2$ powder (Fig. 5) consists of one wide band with local peaks at 435 nm (1.95 eV), 498 nm (2.49 eV), 630 nm (1.97 eV) and 756 nm (1.64 eV). The spectrum of $SnO_2$ nanopowder observed in [28] has the same structure. It can be seen from Fig. 5 that the SPS intensity is comparable to the average intensity of the PCL spectra. A step in the PCL spectrum of $SnO_2$ powder in region of 500 nm can be attributed to the common band in spectra of all the samples measured.

We will assume that the step in the PCL spectrum of $SnO_2$ in the region of 500 nm (2.48 eV) is a manifestation of a separated luminescence band centered in the region of 500 nm.

## 5. Discussion

As seen in Fig. 2-5, the PCL spectra of $SiO_2$, $GeO_2$, $La_2O_3$, $SnO_2$, $TiO_2$ and $CaCO_3$ ultra-pure powders possess a common band centred at 500 nm. The PCL spectra of α-quartz, Ca:$YVO_4$ and $PbWO_4$ crystals and $Y_2O_3$ and $Sc_2O_3$ ultra-pure powders possess a common band centred at 490 nm, and the PCL spectra of $LiNbO_3$ and Sc:$LiNbO_3$ possess a common band centred at 507 nm. We make an attempt to explain these results on the basis of the next assumption: all common bands in the PCL spectra of all the samples studied have an identical origin. Further we will call these bands a "common band".

We suppose that the origin of these bands is related to formation of defects in the samples, which are related to oxygen atoms. This conclusion can be drawn from the fact that the only common feature in all the materials studied is the presence of oxygen.

Firstly, we will discuss the nature of the band with a maximum at 507 nm in the PCL spectra of $LiNbO_3$ and Sc:$LiNbO_3$ crystals (Fig. 2). The doping ion $Sc^{3+}$ replaces $Li^+$ and $Nb^{5+}$ ions in $LiNbO_3$. The valence of $Li^+$ and $Nb^{5+}$ is different to the valence of $Sc^{3+}$. This leads to formation of vacancies to keep the crystal electrically neutral. Thus, the Sc:$LiNbO_3$ crystal possesses a higher concentration of vacancies than the $LiNbO_3$ crystal. Fig. 2 shows that the band with a maximum of intensity at 507 nm is a common band both for the $LiNbO_3$ and Sc:$LiNbO_3$ crystals. However, the relative intensity of a band centred at 507 nm with respect to the intensity of SPS band at 337 nm is ~8 times higher in case of Sc:$LiNbO_3$ in comparison with $LiNbO_3$.

The fact that an increase of the intensity of a band centred at 507 nm is linked to an increase of vacancy concentration leads us to the conclusion that this band is related to vacancies. Further discussion will show that an increase of any vacancy concentration (cationic and aniconic) will lead to the increase of intensity of the related band. According to the assumption of identity of origins of 500 nm, 490 nm and 507 nm PCL bands of $SiO_2$, $GeO_2$, $La_2O_3$, $SnO_2$, $TiO_2$, $CaCO_3$, Ca:$YVO_4$, $PbWO_4$, $Y_2O_3$, $Sc_2O_3$, $LiNbO_3$ and Sc:$LiNbO_3$ samples we suppose that all these bands are related to the vacancies.

Next we will discuss the nature of a band with maximum at 500 nm in the PCL spectrum of $SiO_2$ micro-sized powder (Fig. 2 and 6). The PL band in the spectrum of $SiO_2$ nanopowder with maximum of intensity at 500 nm was attributed in [18-20] to luminescence of OH radicals. We do not attribute the PCL band with maximum of intensity at 500 nm of micrometer-sized powder (Fig. 2) to the luminescence OH-groups, adsorbed on the surface of powder particles, because we took measures to prevent the contamination of samples by water vapour (we used the fresh bags with ultra-pure powders and we annealed the $La_2O_3$ sample at 1100°C before the PCL measurement). Also, it is well-known that OH adsorption leads to an increase in material conductivity [46]. In addition to this, it is important that in our experiment we studied the micro-sized powders. The role of surface effects in comparison with volume effects is less in the case of micro-sized powders in comparison with nano-sized powders.

Let us take a closer look to structure of the PCL spectra of $SiO_2$ powder and α-quartz (Fig. 6). The spectrum of $SiO_2$ powder consists of a flat peak with constant maximum intensity at 300-325 nm, a wide band with maximum at 440 nm, wide band with maximum at 500 nm with a step in region of 520 nm, and a wide band at 635 nm.

The PCL spectrum of an α-quartz crystal consists of two bands. The first band is centered in the region of 415 nm and the second band is centred in the region of 490 nm. The band at 490 nm possesses a step at 523 nm. It is important for the next discussion that the step in 490 nm band of the α-quartz PCL spectrum is positioned at the same wavelength as the step on the 500 nm band of $SiO_2$ powder. This fact is emphasized in Fig. 6 using a vertical line.

We suppose that the coincidence of wavelengths of luminescence bands in the region of 520 nm in the PCL spectra of ultra-pure α-quartz and ultra-pure $SiO_2$ powder is due to the similar nature of these bands in the PCL spectra of α-quartz and $SiO_2$. This proposition can be made because these materials differ from each other only in structural order. Chemical compositions of the samples are identical to a high accuracy. Thus we assume that the luminescence bands in the region of 520 nm in the PCL spectra of an α-quartz crystal and $SiO_2$ powder are associated with the same defect of crystal structure.

The PCL spectrum of α-quartz possesses another band with a maximum of intensity at 420 nm. This band is attributed to STE recombination [2-4,6,8,10-13,16]. It was shown in [20] that the STE band has a red shift with decreasing size of powder particles. Thus, a band with maximum of intensity at 440 nm in the spectrum of $SiO_2$ powder can also be attributed to STE with the appropriate red shift. Also the band in the region of 440 nm can be attributed to an oxygen deficiency center II (ODC II) [26].

Next we will make an assumption that the band with maximum at 490 nm in the PCL spectrum of an α-quartz crystal and the band with maximum at 500 nm in the spectrum of $SiO_2$ powder have an identical nature. The difference at the positions of maxima of these bands may be explained by the different admixture of the band at 520 nm from the intensity of the band at 490 nm in the case of crystal and powder. The intensity of the band at 520 nm is very low in the case of the α-quartz crystal, unlike in the case of $SiO_2$ powder where the intensity is significant and can provide a perceptible shift of maximal intensity of the band in the region of 490-500 nm.

Taking into account these assumptions it can be seen that all the PCL bands of an α-quartz crystal are contained in the PCL spectrum of $SiO_2$ powder. At the same time only the PCL spectrum of $SiO_2$ powder contains the bands with maxima at 635 nm and in the range of 300-325 nm.

The band with a maximum at ~3 eV in the emission spectrum of pure crystalline quartz was observed in [10] under the excitation by X-rays and had been attributed to STE recombination. However, the band with a maximum at 500 nm (~2.5 eV) had not been observed in [10]. We suppose that the absence of a ~2.5 eV band in the case of X-ray excitation is due to a high penetration depth of X-radiation in quartz in comparison with the electron beam penetration depth with the average energy of particles ~150 keV. The electron beam excites the luminescence in a thin surface layer of the sample with ~0.1 mm thickness. In this layer the concentration of vacancies is higher than in a volume of the sample because of the proximity of the crystal surface. So, the relative intensity of defect luminescence in comparison with STE luminescence is negligible in the case of X-ray excitation.

Defects in $SiO_2$ can be divided into two groups [26]. The first group consists of defects, the properties of which are determined in the first order by interactions within a single $SiO_4$ tetrahedron. NBOHC, POR (Peroxy Radical), dangling-bond defects, 3-fold and 2-fold-coordinated Si atoms belong to this group. The second group consists of defects the properties of which are determined, in the first order, by interaction between $SiO_4$ tetrahedrons. This group consists of the oxygen vacancy with a trapped hole (E'-centre), a relaxed diamagnetic oxygen vacancy, oxygen divacancies and 'non-relaxed' oxygen vacancies. An existence of unrelaxed diamagnetic oxygen vacancy is not clear yet [26]. Thus, we will imply the relaxed oxygen vacancy by the term 'vacancy'.

In [26] it was reported that defects of the first group cannot occur in α-quartz due to the steric limitations. Defects of the first group can appear in α-quartz only after the irradiation by the massive particles (neutrons [17,47], protons, etc.). According to [26], defects of first type have never been observed in crystalline quartz single crystals irradiated by γ-rays. This fact leads to the conclusion that the existence of dangling-bond defects in crystalline quartz is impossible without shifting of the nuclear positions.

Amorphization of α-quartz under the action of a highly focused electron beam with fluence $10^{23}$-$10^{27}$ e/m$^2$ (e denotes an electron) with particles energy of 200 keV is observed in [48]. This fluence corresponds to the adsorption by a quartz sample of the energy with spatial density ~$10^3$-$10^7$ J/cm$^2$. Spatial energy density of the RADAN-EXPERT accelerator, integrated over the one pulse, is ~$3\cdot10^{-2}$ J/cm$^2$ [35]. Thus we can neglect the amorphization of α-quartz during the PCL measurements.

Thus, the origins of the band at 250-320 nm and a band with maximum of intensity at the region of 635 nm in the PCL spectrum of $SiO_2$ powder are related to the defects of the first group. This is consistent with [4,7,14,15,18-20,26], where the band at 1.9 eV of amorphous $SiO_2$ is related to the NBOHC. Thus, the band with maximum of intensity at 500 nm in the PCL spectrum of $SiO_2$ powder and band with maximum of intensity at 490 nm in the PCL spectrum of α-quartz both should be attributed to the defects of the second group.

Let us consider the neutral oxygen vacancy in α-quartz [49-54]. This defect belongs to the second group. The distance between two silicon atoms in the absence of an oxygen atom between them is calculated in [50,51] to be equal to 2.52 Å in comparison with the case of absence of the oxygen vacancy, when the distance is 3.06 Å. It is seen that the oxygen vacancy leads to a significant stress in the crystalline lattice. This stress leads to a neighbouring tetrahedron violation. Thus, each vacancy possesses a number of neighbouring violated tetrahedrons. These tetrahedrons have no dangling bonds. However, the shape and orientation of these tetrahedrons are slightly violated in comparison with the tetrahedrons from a perfect lattice. It is important to notice that not only an oxygen vacancy can lead to stress in the lattice. Stress in the lattice can be related to any type of vacancy (cationic or anionic).

Luminescence of neutral vacancy-related oxygen deficiency centers is not, to our knowledge, clearly understood yet [26], because this diamagnetic defect is invisible for the electron paramagnetic resonance (EPR) method [50]. At the same time, a neutral vacancy is described as a probable precursor for the EPR-active E'-center [26,49-54].

We propose the next model of the defect which corresponds to the emission at 500 nm in the PCL spectrum of $SiO_2$ powder and at 490 nm in the PCL spectrum of crystalline quartz. Two neighbouring $SiO_4$ tetrahedrons with violated geometry and orientation possess two $O^{2-}$ ions which are closer to each other in comparison with the perfect crystal lattice. This symmetry violation leads to the convergence of these two oxygen ions via the Coulomb attraction, when one of them is ionized during the PCL excitation. This defect can arise in each of the oxide samples studied and becomes insensitive in the first order to the crystal environment. Thus, we attribute the luminescence at 500 nm of $SiO_2$ powder and luminescence at 490 nm of α-quartz to the recombination of $O^{2}$-$O^{-}$ complex at the region of tetrahedron $SiO_4$ violation.

We suppose that the observed bands centred at 490 nm in the PCL spectrum of crystalline quartz and at 500 nm in the PCL spectrum of $SiO_2$ powder are not related to the neutral oxygen vacancy itself. According to [35], the energy of this defect is related directly to Si-Si bond energy. Dependence of the emission wavelength on the Si-Si bond is in contradiction with our assumption that the bands at 500 nm in the PCL spectra of $SiO_2$, $GeO_2$, $La_2O_3$, $TiO_2$ and $CaCO_3$ powders (Fig. 2) has the same nature. For example, the bond dissociation energy of the Ge-Ge bond is ~264 kJ/mol, Ti-Ti bond is ~118 kJ/mol, La-La bond is ~245 kJ/mol and C-C bond is ~618 kJ/mol [54]. Independence of the position of the band at 500 nm in $SiO_2$, $GeO_2$, $La_2O_3$, $TiO_2$ and $CaCO_3$ powders on the chemical composition proves that this defect is not directly related to Si, Ge, La, Ti and C atoms and has to be related to the oxygen atoms. At the same time, the analysis of the PCL spectra of $LiNbO_3$ and $Sc:LiNbO_3$ crystals (see Fig. 3) leads to the conclusion that an increase of vacancy concentration leads to the increase of the intensity of a band at 507 nm. We suppose this band has the same origin as the 500 nm band in the PCL spectra of $SiO_2$, $GeO_2$, $La_2O_3$, $TiO_2$ and $CaCO_3$ powders. Thus, this defect has to be related to the interaction between the neighbouring violated coordination polyhedrons of these materials in the vicinity of vacancies.

The coordination polyhedron in the case of $CaCO_3$ is a $CO_3$ triangle, in the case of $TiO_2$ and $GeO_2$ –$TiO_6$ and $GeO_6$ octahedral and in the case of $La_2O_3$ – $LaO_7$ polyhedron. Independence of the PCL peak position for these materials on the structure of the coordination polyhedron additionally supports the assumption that a band centred at 500 nm is formed due to the defects related to the interaction between different coordination polyhedrons.

The PCL spectrum of $La_2O_3$ powder like the PCL spectrum of $SiO_2$ powder contains a band in the region of 630 nm (Fig. 2). PCL bands in the region of 500-800 nm have not been observed in the spectra of $TiO_2$, $GeO_2$ and $CaCO_3$ ultra-pure powders. We attribute this result to the low symmetry of $SiO_4$ tetrahedrons and $LaO_7$ polyhedrons in comparison with the higher-symmetrical octahedrons $TiO_6$ and $GeO_6$ and $CO_3$ triangle. We suppose that a higher symmetry of the polyhedron leads to the denser arrangement of polyhedrons and provides more possibilities for the steric limitation on the defects inside the coordination polyhedron.

The PCL spectra of $Ca:YVO_4$ and $PbWO_4$ crystals and $Y_2O_3$ and $Sc_2O_3$ powders are shown in Fig. 3-4. It is seen that the common band in the PCL spectra of these materials is situated at 490 nm. According to the previous discussion we attribute this band to recombination of $O^{2-}$-$O^-$ complexes in the regions of a coordination polyhedron violation. This band is dominant in the PCL spectrum of $Ca:YVO_4$ (Fig. 3). This result supports our assumption because the $Ca^{2+}$ doping leads to oxygen vacancy formation which increases the number of regions with coordination polyhedron violation.

A common feature of $Ca:YVO_4$, $PbWO_4$, $Y_2O_3$ and $Sc_2O_3$, in comparison to the $SiO_2$, $GeO_2$, $TiO_2$, $CaCO_3$, $LiNbO_3$ and $Sc:LiNbO_3$ materials is a direct band gap transition. Thus, we suppose that the emission wavelength of the discussed common band is ~490 nm. Red shift in the case of the $SiO_2$, $GeO_2$, $La_2O_3$, $TiO_2$, $CaCO_3$, $LiNbO_3$ and $Sc:LiNbO_3$ materials is due to the indirect band gap transition: a difference of energy is due to the phonon emission. However, crystalline α-quartz is an indirect solid [8,55]. It will be shown further that the phonon emission in case of α-quartz is not necessary during the recombination of $O^{2-}$-$O^{-}$ complexes due to the fact that the bottom of the conduction band in α-quartz is situated at the Γ point of the Brillouin zone.

A manifestation of the direct character of band gap transition of Ca:$YVO_4$, $PbWO_4$, $Y_2O_3$ and $Sc_2O_3$ materials is also a high average intensity of PCL spectra in comparison with the α-quartz, $SiO_2$, $GeO_2$, $La_2O_3$, $TiO_2$, $CaCO_3$, $LiNbO_3$ and Sc:$LiNbO_3$. This fact follows from the comparison of intensities of SPS emission and luminescence of the material (Fig. 3-4).

The structure of the PCL spectrum of $SnO_2$ powder (Fig. 5) is different from the structure of the PCL spectra of all other samples studied. It consists of one wide band with local maxima of intensity at 435 nm, 498 nm, 630 nm and 756 nm. This result is in agreement with [28]. The luminescence of $SnO_2$ nanopowder is attributed there to radiative decay of electrons trapped at oxygen vacancies, and from the conduction band to intrinsic surface states of the powder particles. It is important for our discussion that the PCL spectrum of $SnO_2$ powder possesses a step in the region of 500 nm. As previously, we also attribute this luminescence to recombination of $O^{2-}$-$O^{-}$ complexes in the regions of the $SnO_6$ octahedron violation.

Structural difference between the PCL spectrum of $SnO_2$ powder and spectra of other samples studied can be explained by the weakest chemical bonding of Sn-O between cation and anion in comparison with other materials studied (see Table 1). This fact leads to the presence of the largest amount of broken chemical bonds on the surface of $SnO_2$ powder particles in comparison with other samples studied. This leads to the formation of a large number of surface states. Thus, the luminescence is determined primarily by these states. This fact had been proven in [28]. The position of a band at 500 nm is in line with the previous discussion, because the intensity of PCL of $SnO_2$ powder is comparable to the intensity of SPS bands (Fig. 5). These facts are consistent with the indirect character of band gap transition of $SnO_2$ (see Table 1).

As can be seen from the previous discussion that position of maximum intensity of a common band depends on the type of band gap transition of the material. This fact can be explained as follows: the $O^{2-}$-$O^{-}$ defect is spatially localized due to the absence of this defect in other points of sample apart from in the vicinity of vacancies. Thus, this defect cannot tunnel to other points of sample and its momentum k is extremely small: k~0. Thus, it is impossible for this defect to recombine with an electron from the conduction band with non-zero momentum without a phonon emission. It is seen that indirect materials should have a weak luminescence at the wavelength of a common band. It can be seen from above, Fig. 2-6, and Table 1, that this statement is fully justified.

In the case of crystalline α-quartz, the top of the valence band is situated at the K point of the Brillouin zone, according to [55]. The bottom of the conduction band is situated at the Γ point. Thus, α-quartz is an indirect solid. At the same time, the position of the common band in the PCL spectrum of crystalline α-quartz corresponds to the case of material with a direct band gap transition. The Γ point of the Brillouin zone corresponds to zero quasi-momentum [57]. This fact leads to the situation when the recombination of a defect does not require a phonon emission and the wavelength of the common band is 490 nm despite the fact that crystalline α-quartz is an indirect solid.

It can be seen from Fig. 2-6, that the intensity of the common band is equal in order of magnitude to the intensity of luminescence of other structural defects of the materials. This fact makes it possible to use the common band as a universal spectroscopic marker of ultra-pure oxides.

Summarizing, we introduce a common model of PCL of oxides with a low concentration of optically-active impurities. All measured solid oxides possess a PCL band at 490-507 nm. We suppose this band to have a common origin in PCL spectra of all the samples studied. This band is a common feature of oxide materials. It has no relation to chemical composition, chemical bond ionicity, coordination polyhedron type or a crystal order. This band is related to an $O^{2-}$-$O^{-}$ complex which appears during the luminescence excitation in the regions of coordination polyhedron geometry and orientation violation. These regions exist in the vicinities of anionic and cationic vacancies due to the stress in the lattice. This band is a universal marker of a high-purity oxide.

## 6. Conclusion

Pulsed Cathodoluminescence (PCL) spectra of ultra-pure $SiO_2$, $GeO_2$, $SnO_2$, $TiO_2$, $La_2O_3$, $Y_2O_3$, $Sc_2O_3$, $CaCO_3$ powders and α-quartz, Ca:$YVO_4$, $LiNbO_3$ and Sc:$LiNbO_3$ crystals were studied under the same experimental conditions. All materials had a low concentration of optically-active impurities.

It was found that the PCL spectra of $SiO_2$, $GeO_2$, $TiO_2$, $SnO_2$, $La_2O_3$ and $CaCO_3$ powders contain a common band with a maximum intensity at 500 nm, the PCL spectra of samples $Y_2O_3$, $Sc_2O_3$, $PbWO_4$ and Ca:$YVO_4$ contain a common band at 490 nm and the PCL spectra of $LiNbO_3$ and Sc:$LiNbO_3$ crystals contain a common band at 507 nm.

The position of these bands is not affected by the chemical composition, structural order, percentage of ionicity of chemical bonding or type of coordination polyhedron of the material. It is found, however, that the average intensity of the PCL spectra and position of maximal intensity of these common bands depend on the type of a band gap transition of the material.

Materials with a direct band gap possess a common band in the PCL spectra in the region of 490 nm. Materials with an indirect band gap possess a red-shifted common band due to the phonon emission during the luminescence transition.

We suppose that these common bands in all the materials studied have the same origin. We attribute these common bands to the luminescence of $O^{2-}$-$O^{-}$ oxygen complexes. These complexes appear in the vicinities of anionic or cationic vacancies, where the geometry and orientation of coordination polyhedrons are violated due to the stress in the crystalline lattice.

Presence of this common band in the PCL spectra of oxides is a marker of a low concentration of optically-active impurities.

**Acknowledgements**

This work is supported by the Russian Science Foundation (project No. 14-22-00273).

**Table 1.**

Characteristics of samples studied

| Chemical Formula of Material | Type of Coordination Polyhedron | Band Gap Width, eV | Type of Band Gap Transition | Chemical Bonding to Oxygen | Bond Dissociation Energy (kJ/mol) [54] |
|---|---|---|---|---|---|
| $SiO_2$ | $SiO_4$ tetrahedron | 11.2 [11] | indirect (α-$SiO_2$) [55] | Si-O | 799 |
| $TiO_2$ | $TiO_6$ octahedron | 3, 3.3, 3.4 [68] | indirect [67] | Ti-O | 666 |
| $GeO_2$ | $GeO_4$ tetrahedron<br>$GeO_6$ octahedron | 5.7 [70] | indirect (α-$GeO_2$) [71] | Ge-O | 657 |
| $SnO_2$ | $SnO_6$ octahedron | 3.8 [59] | indirect [69] | Sn-O | 528 |
| $La_2O_3$ | $LaO_7$ polyhedron | 5.5 [64] | no data | La-O | 798 |
| $Y_2O_3$ | $YO_6$ octahedron | 6.2 [59] | direct [58,65] | Y-O | 714 |
| $Sc_2O_3$ | $ScO_6$ octahedron | 6.0 [59] | direct [65] | Sc-O | 671 |
| $CaCO_3$ | $CO_3$ triangle | 6.0 [63] | indirect [62] | C-O | 1076 |
| Ca:$YVO_4$ | $VO_6$ octahedron | 3.8 [62] | direct [62] | V-O | 637 |
| $PbWO_4$ | $WO_6$ octahedron | 4.2 [61] | direct [61] | W-O | 720 |
| $LiNbO_3$ | $NbO_6$ octahedron<br>$LiO_3$ triangle | 3.6 [60] | indirect [60] | Nb-O<br>Li-O | 726<br>340 |
| Sc:LiNbO3 | $NbO_6$ octahedron<br>$ScO_6$ octahedron<br>$LiO_3$ triangle | 3.6 [60] | indirect [60] | Nb-O<br>Li-O<br>Sc-O | 726<br>340<br>671 |

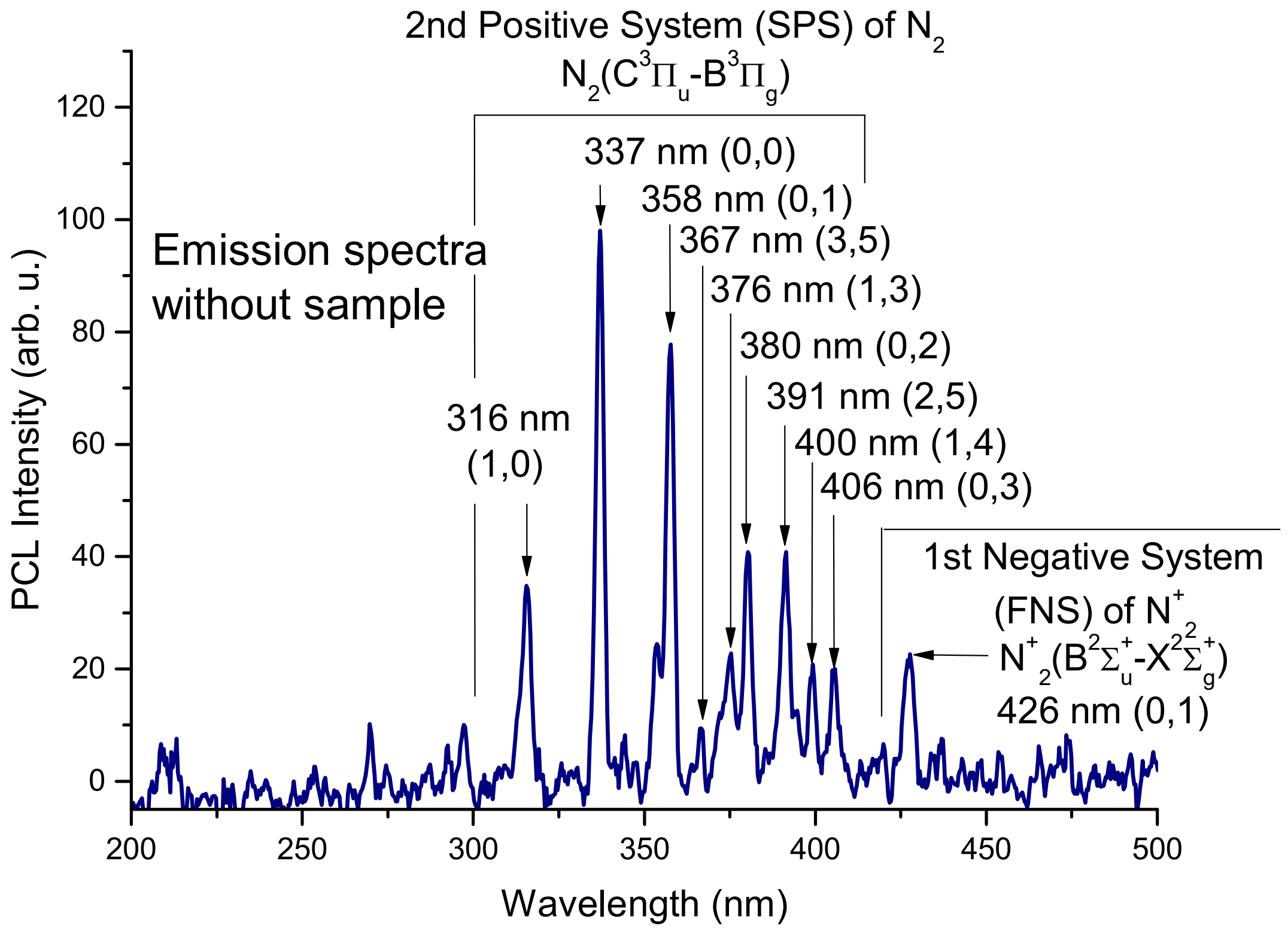

**Fig. 1**. Emission spectra of ambient air, excited by electron beam with average energy of particles ~150 keV. Spectra are attributed to the second positive system (SPS) of molecular nitrogen $N_2$ and first negative system (FNS) of molecular ion $N_2^+$ [41]. The wavelength of maxima of intensities and appropriate vibration transitions in breaks are shown.

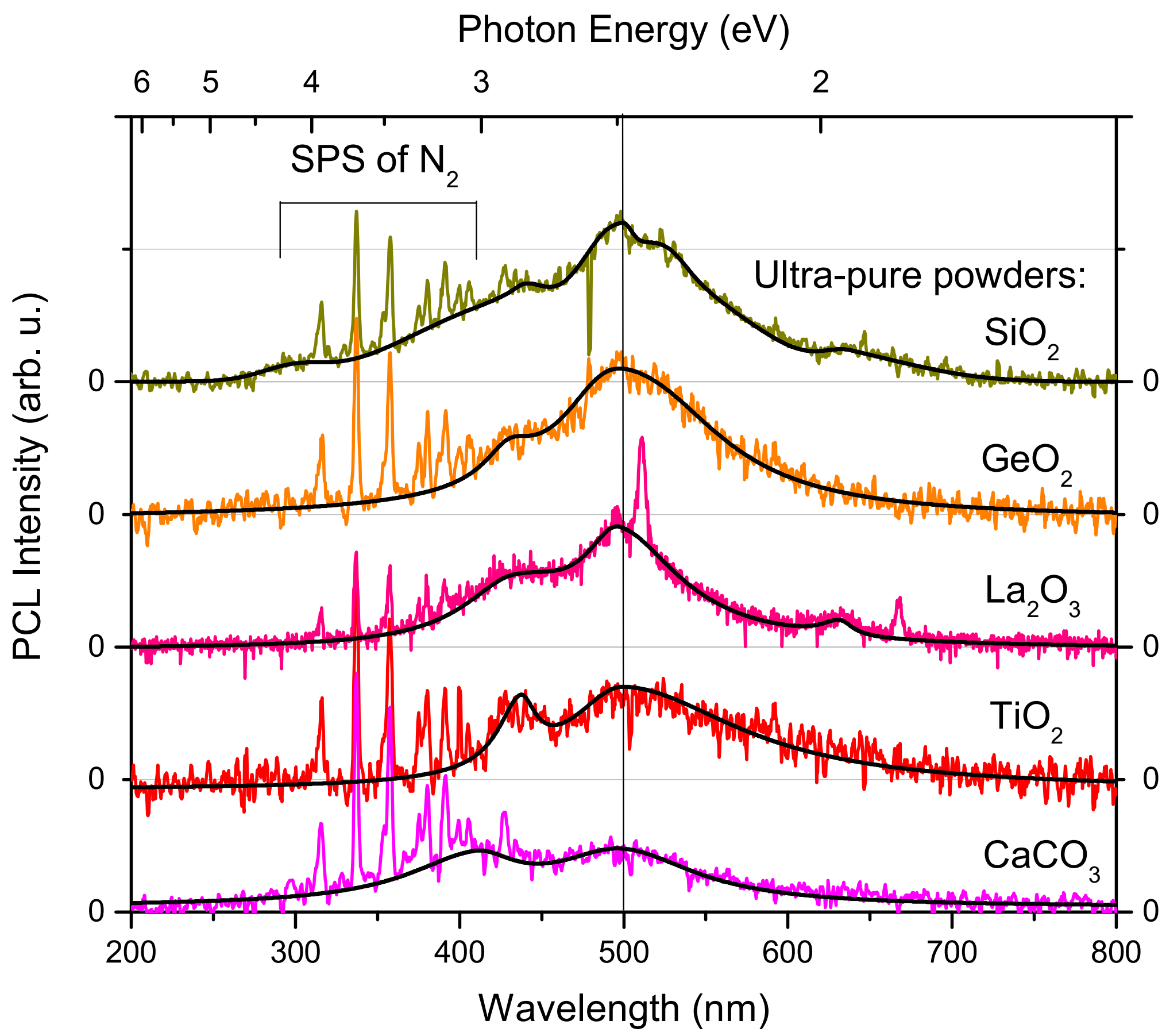

**Fig. 2**. Room-temperature PCL spectra of $SiO_2$, $GeO_2$, $La_2O_3$, $TiO_2$ and $CaCO_3$ ultra-pure powders. The spectra are the result of averaging of 30 pulses. Solid lines show the fitted spectra. The vertical line shows the position of 500 nm.

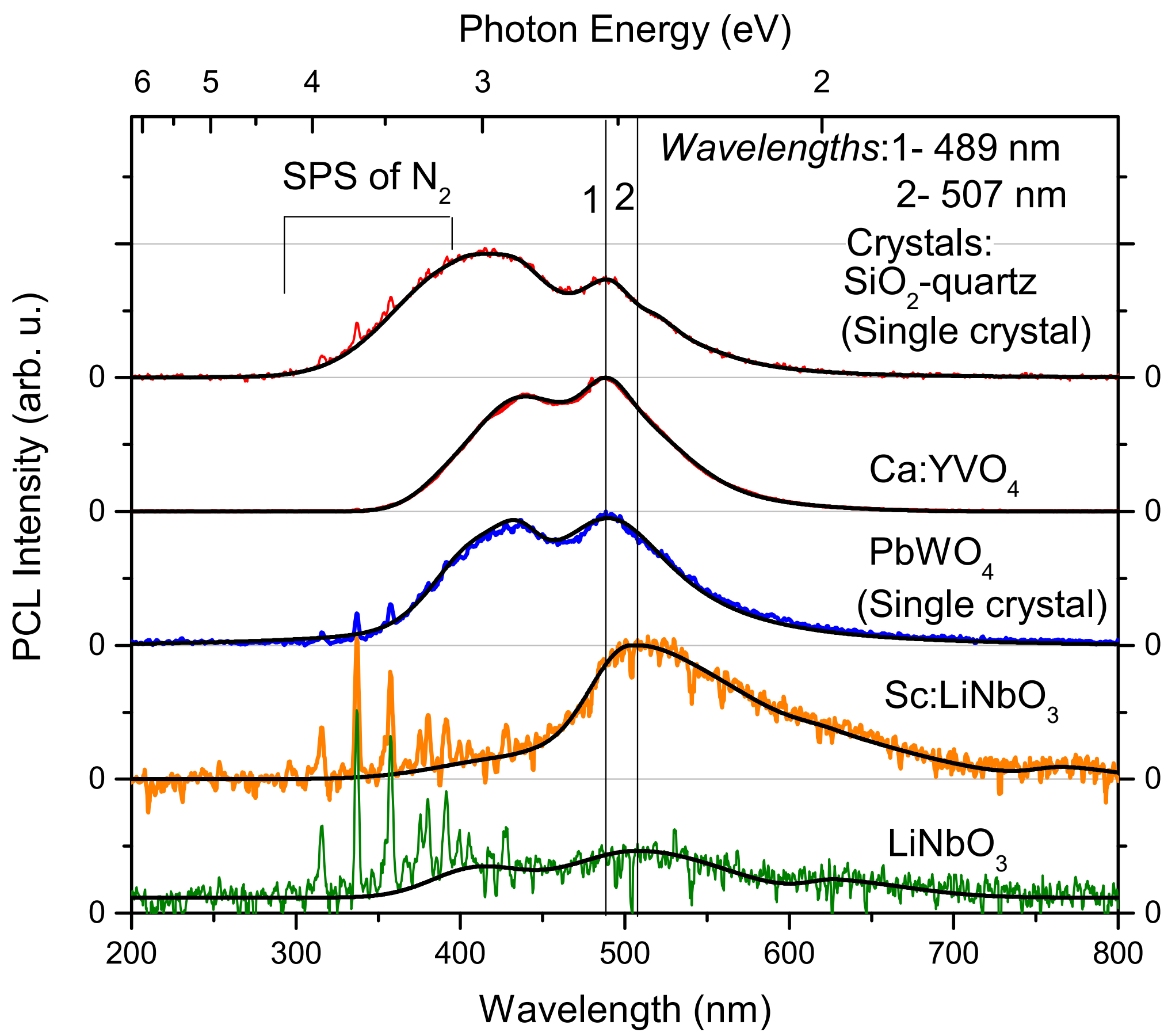

**Fig. 3**. Room-temperature PCL spectra of α-quartz, $Ca:YVO_4$, $PbWO_4$, $LiNbO_3$ and $Sc:LiNbO_3$ crystals. The spectra are the result of averaging of 30 pulses. Solid lines show the fitted spectra. The vertical lines correspond to wavelengths of 489 nm (1) and 507 nm (2).

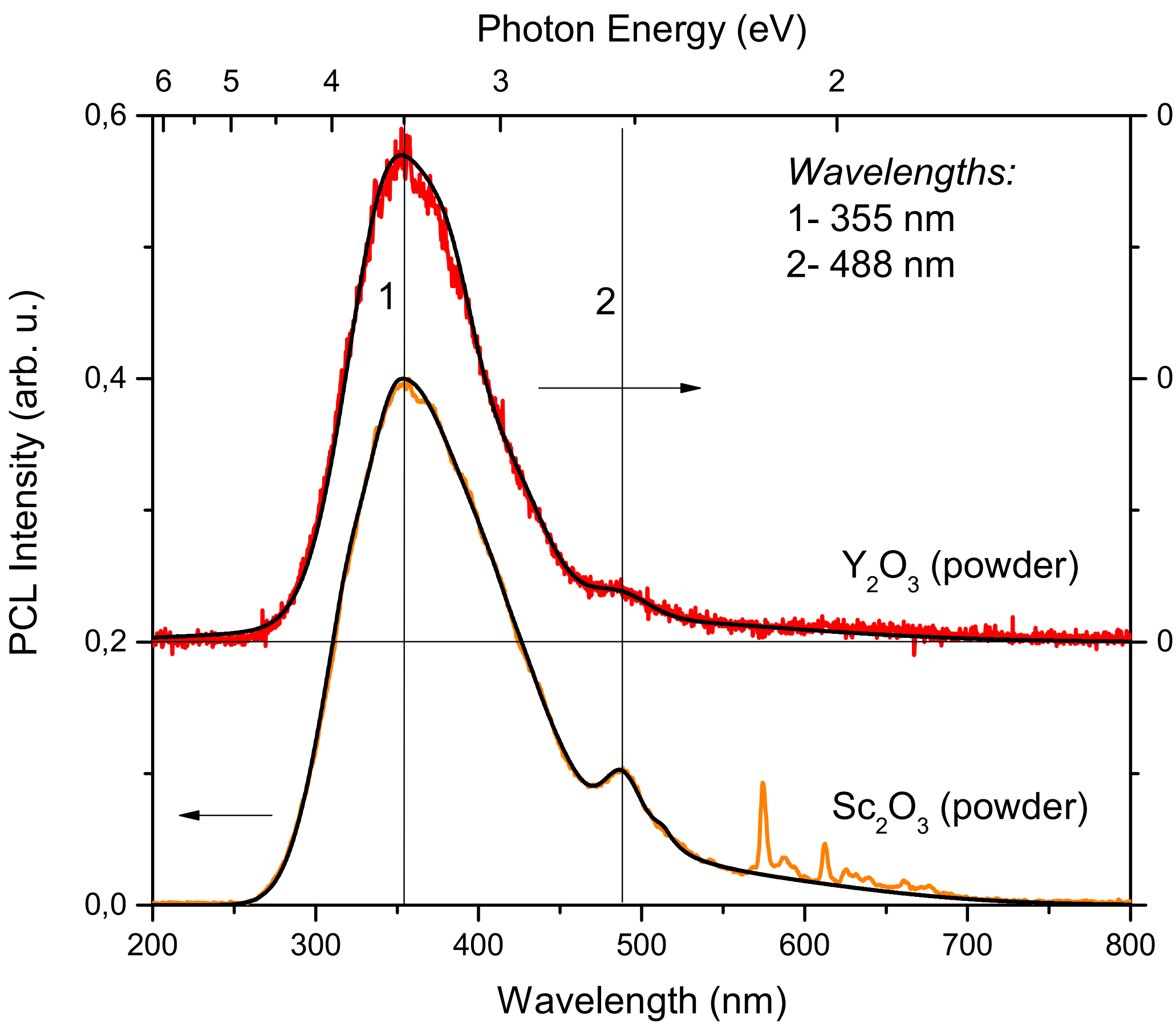

**Fig. 4**. Room-temperature PCL spectra of $Y_2O_3$ and $Sc_2O_3$ ultra-pure powders. The spectra are the result of averaging of 30 pulses. Solid lines show the fitted spectra. The vertical lines correspond to wavelengths of 355 nm (1) and 488 nm (2). The rows show the corresponding vertical axes.

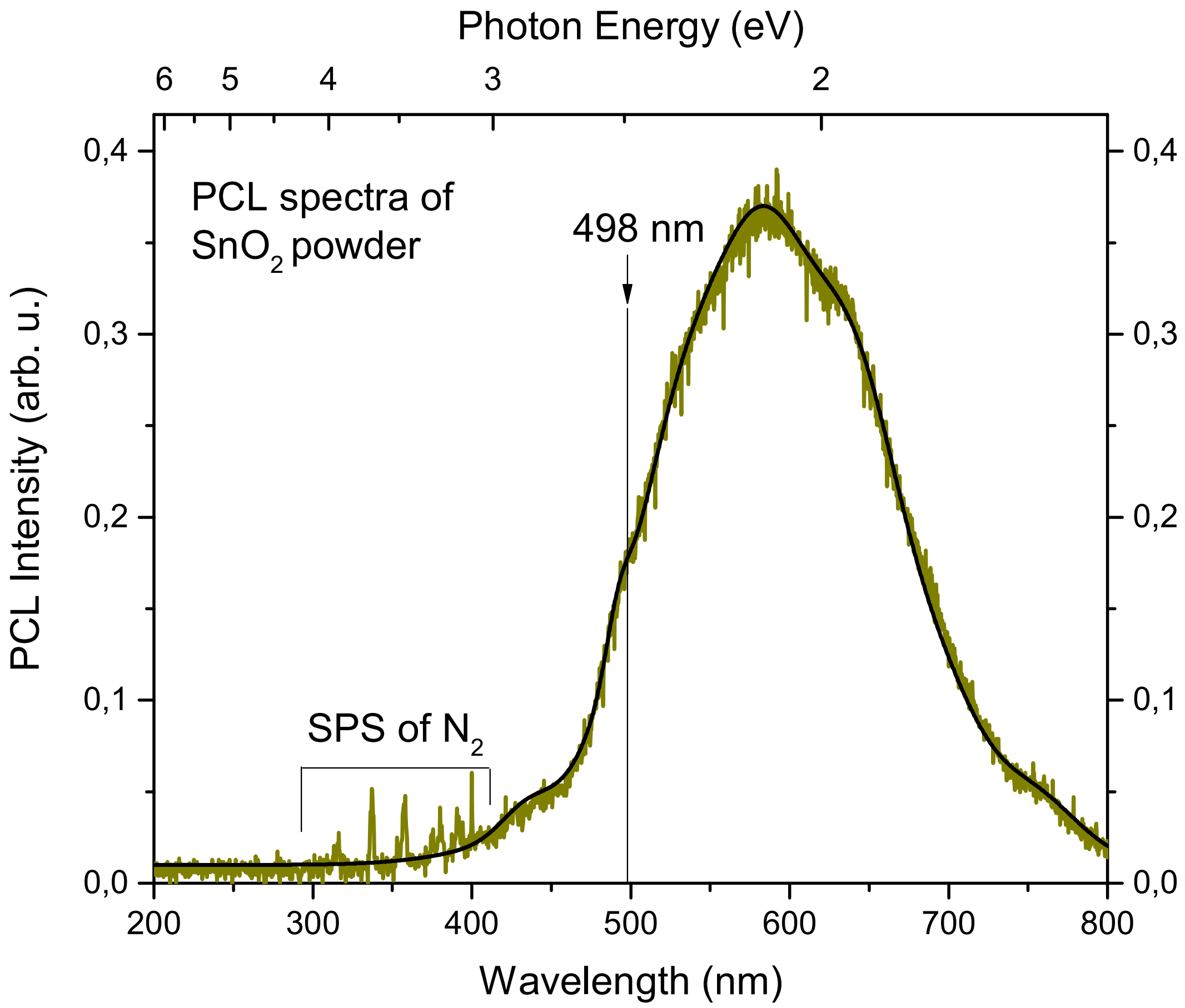

**Fig. 5**. Room-temperature PCL spectra of $SnO_2$ ultra-pure powder. The spectra are the result of averaging of 30 pulses. The solid line shows the fitted spectra.

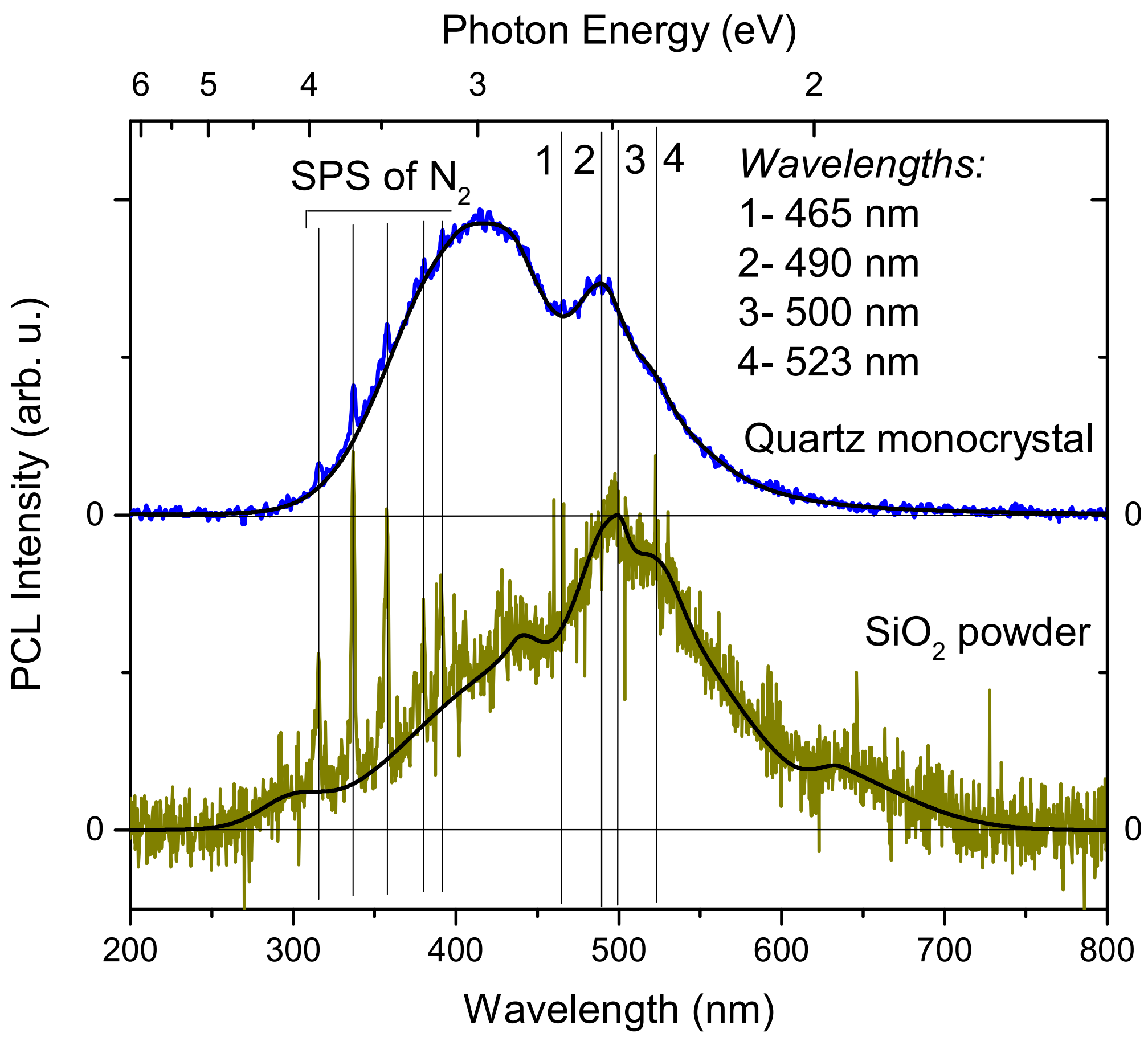

**Fig. 6**. Comparison of room-temperature PCL spectra of ultra-pure hydrothermal α-quartz and ultra-pure $SiO_2$ powder. Solid lines show the fitted spectra. The vertical lines correspond to wavelengths of 465 nm (1), 490 nm (2), 500 nm (3) and 523 nm (4).